\documentclass[aps,prx,twocolumn,nofootinbib,superscriptaddress,longbibliography,notitlepage,floatfix,10pt]{revtex4-2}
\usepackage{graphicx,outlines,subfigure}
\usepackage{physics,graphicx,diagbox,markdown}
\usepackage{tabularx}
\usepackage{amsmath,mathtools}
\usepackage{amstext,cuted}
\usepackage{amssymb}
\usepackage{xfrac}
\usepackage[colorlinks,citecolor=magenta]{hyperref}
\usepackage{graphicx,makecell}
\usepackage{amsmath}
\usepackage{amstext}
\usepackage{amssymb}
\usepackage{longtable,booktabs}
\usepackage{hyperref}
\usepackage{url}
\usepackage{subfigure}
\usepackage{dsfont}
\usepackage{booktabs}
\usepackage{amsbsy}
\usepackage{dcolumn}
\usepackage{amsthm}
\usepackage{times}
\usepackage{bm}
\usepackage{esint}
\usepackage{multirow}
\usepackage{hyperref}
\usepackage{cleveref}
\usepackage{amsmath}
\usepackage{amssymb}
\usepackage{mathrsfs}
\usepackage{amsbsy}
\usepackage{dcolumn}
\usepackage{bm}
\usepackage{multirow}
\usepackage{color}
\usepackage{extarrows}
\usepackage{datetime}
\usepackage{comment}
\usepackage[super]{nth}
\usepackage{tikz}
\usetikzlibrary{arrows.meta}
\hypersetup{
	colorlinks=true,
	linkcolor=blue,
	filecolor=blue,
	urlcolor=blue,
}

\usepackage{datetime}
\usepackage{comment}
\usepackage{lineno}
\graphicspath{{Pictures/}}
\usepackage{xcolor}

\begin{document}

\preprint{APS/123-QED}

\title{Entanglement scaling behaviors of free fermions on hyperbolic lattices}

\author{Xiang-You Huang}
\thanks{These authors contributed equally to this work.}
\affiliation{Guangdong Provincial Key Laboratory of Magnetoelectric Physics and Devices, State Key Laboratory of Optoelectronic Materials and Tecnologies,
	and School of Physics, Sun Yat-sen University, Guangzhou, 510275, China}

\author{Yao Zhou}\thanks{These authors contributed equally to this work.}
\affiliation{Guangdong Provincial Key Laboratory of Magnetoelectric Physics and Devices, State Key Laboratory of Optoelectronic Materials and Tecnologies,
	and School of Physics, Sun Yat-sen University, Guangzhou, 510275, China}
\affiliation{Department of Physics, The University of Hong Kong, Pokfulam Road, Hong Kong SAR, China}

\author{Peng Ye}

\email{yepeng5@mail.sysu.edu.cn}

\affiliation{Guangdong Provincial Key Laboratory of Magnetoelectric Physics and Devices, State Key Laboratory of Optoelectronic Materials and Tecnologies,
	and School of Physics, Sun Yat-sen University, Guangzhou, 510275, China}

\date{\today}

\begin{abstract}
Recently, tight-binding models on hyperbolic lattices (discretized AdS space) have gained significant attention, leading to hyperbolic band theory and non-Abelian Bloch states. In this paper, we investigate these quantum systems from the perspective of quantum information, focusing particularly on the scaling of entanglement entropy (EE) that has been regarded as a powerful quantum-information probe into exotic phases of matter. It is known that on $d$-dimensional translation-invariant Euclidean lattice, the EE of band insulators scales as an area law ($\sim L^{d-1}$; $L$ is the linear size of the boundary between two subsystems). Meanwhile, the EE of metals (with finite Density-of-State, i.e., DOS) scales as the renowned Gioev-Klich-Widom scaling law ($\sim L^{d-1}\log L$). The appearance of logarithmic divergence, as well as the analytic form of the coefficient $c$ is mathematically controlled by the Widom conjecture of asymptotic behavior of Toeplitz matrices and can be physically understood via the Swingle's argument. However, the hyperbolic lattice, which generalizes translational symmetry, results in inapplicability of these analytic approaches and the potential non-trivial behavior of EE. Here we make an initial attempt through numerical simulation. Remarkably, we find that both cases adhere to the area law, indicating the effect of background hyperbolic geometry that influences quantum entanglement. To achieve the results, we first apply the vertex inflation method to generate hyperbolic lattice on the Poincar\'{e} disk, and then apply the Haydock recursion method to compute DOS. Finally, we study the scaling of EE for different bipartitions via exact diagonalization and perform finite-size scaling. We also investigate how the coefficient of the area law is correlated to bulk gap in gapped case and to the DOS in gapless case respectively. Future directions are discussed.
\end{abstract}

\maketitle


\section{Introduction}
\label{sec:introduction}

Quantum information theory provides a novel approach to study non-local correlations of quantum many-body systems~\cite{Amico_2008_review,Eisert_2010_review,Laflorencie_2016_review}. To quantify these non-local correlations, the celebrated entanglement entropy (EE, or von Neumann entropy) plays an important role and exhibits universal features. 
For instance, the scaling behavior of EE reveals the underlying nature of the systems~\cite{Amico_2008_review,Eisert_2010_review,Laflorencie_2016_review,Calabrese_2004,Abanin_2019_c,Haas_2006_S,Barthel_2006_E,Zhou_2023_E}.
In systems with energy gap, the leading term of EE for ground states satisfies the area-law $S_A \sim L_A^{d-1}$~\cite{Eisert_2010_review,Laflorencie_2016_review,Haas_2006_S}, where $d$ is the spatial dimension and $L_A$ is the linear size of the boundary between two complementary subsystems denoted as $A$ and $B$. 
For gapless systems, conformal field theory (CFT) provides an insight into the scaling of EE in $1$d gapless systems~\cite{Holzhey_1994,Calabrese_2009}. Furthermore, for higher-dimensional free-fermion systems with codimension-1 Fermi surface, the application of the Widom conjecture~\cite{Widom1982} gives the scaling of leading term of EE, which leads to  the Gioev-Klich-Widom scaling (also dubbed ``super-area law'') $S_A \sim L_A^{d-1} \log L_A$~\cite{Gioev_2006,Leschke_2014}. Meanwhile, Swingle proposed simple reconstruction method to physically understand the origin of  logarithmic divergence term and the analytic form of the coefficient $c$~\cite{Swingle_2010_E}. 
The logarithmic divergence, to some extent, indicates that the presence of infinite number of gapless fermion modes significantly enhances entanglement. 
 
It is worth noting that these  scaling behaviors  are established on the translation-invariant lattices with Euclidean geometry, where the Widom conjecture of asymptotic behavior of Toeplitz matrices is applicable. 
Therefore, we raise the question whether the behaviors of EE could be significantly changed by the background geometry, as we find that entanglement on fractal lattice can exhibit fractal-like distribution and generalized area law reflecting boundary Hausdorff dimension~\cite{zhou2024quantum}. In fact, non-Euclidean geometry is prevalent in natural and artificial systems~\cite{Magnus1974a}. Anti-de Sitter (AdS) space, characterized by negative spatial curvature, is widely studied in various fields of physics~\cite{witten,Klebanov1998,RT,maldacenaLarge1998,Viswanathan1998a,Viswanathan1998b,Henningson_1998,nishiokaEntanglement2018,Cirac_2021_M}. The hyperbolic lattice, which can be viewed as a discretization of AdS space, is of interest in high energy physics~\cite{Brower_2021_L,Asaduzzaman_2020_H,Brower_2022_H,dey2024simulating}. Recently, hyperbolic lattice has been experimentally simulated on many platforms~\cite{Kollr2019,Yu_2020_T,Zhang2022,Lenggenhager2022,chen_2023_adscft,Huang2024} and draws more and more attentions in various fields of condensed matter physics~\cite{Gendiar_2020_area,Nishino_2024_holographic,2024arXiv240917235S,Zhu_2021,Boettcher_2020_Q,Tummuru_2024_Semimetal,Chen2024anderson,Liu_2022_C,Urwyler_2022_H,Liu_2023_H,Tao_2023_H,Manna_2023_mass,Chen_2023_S,Lux_2023_C,sun2024topological,Lenggenhager2024spin,Dusel2024spin,He2024fci}.  
Hyperbolic lattice is highly different from its Euclidean counterpart due to its non-Abelian translation symmetry~\cite{Boettcher_2022_C,Maciejko_2021_HBT,Maciejko_2022_A,Lenggenhager_2023_N}. Remarkably, these geometric properties lead to the hyperbolic band theory (HBT) for tight-binding models on hyperbolic lattices~\cite{Maciejko_2021_HBT,Maciejko_2022_A,Lenggenhager_2023_N,Cheng2022band,shankar2023hyperbolic}.

The absence of Euclidean translation invariance on hyperbolic lattice results in the inapplicability of Widom conjecture of Toeplitz matrices, implying that the EE may exhibit non-trivial behaviors. 
Motivated by the rapid progress on hyperbolic lattices as well as application of quantum information in many-body physics,  we explore the potential role of hyperbolic geometry in affecting quantum entanglement in this paper. 
However, the analytic difficulties are significantly challenging as the Widom conjecture of Toeplitz matrices is no longer applicable. Therefore, our goal is to  provide numerical evidence of the exotic interplay of quantum entanglement and hyperbolic geometry  by   investigating the scaling of EE of free-fermion systems on hyperbolic lattices. 
We observe that for gapped systems, the EE still scales as the area law, consistent with our expectations on Euclidean lattice. However, for gapless system with finite DOS, we discover that the super-area law breaks down, and the EE adheres to the area law instead. 
This area law scaling reflects non-trivial effect of hyperbolic geometry for entanglement, which may relates to a holographic understanding that could be experimentally studied~\cite{RT,chen_2023_adscft}. 
Moreover, similar to our previous work in fractal geometry as summarized in Table.~\ref{tab:eegeometry}, our results suggest a perspective to study the geometry of quantum systems through entanglement. 
\begin{table}
\caption{\label{tab:eegeometry} Scaling behavior of EE of ground states of free fermionic systems on $d$-\textit{dimensional} Euclidean lattice with translation invariance, fractal lattice with self-similarity and \textit{two-dimensional} hyperbolic lattice. In fractal case, $d_{\rm bf}$ denotes the \textit{Hausdorff dimension} of boundary of subsystem $A$, while EE of gapless systems exhibits fractal-like distribution~\cite{zhou2024quantum}.
}
\begin{ruledtabular}
\begin{tabular}{ccc}
\multirow{2}{*}{Lattice} & \multirow{2}{*}{Phase} & \multirow{2}{*}{$S_{A}$}      \\ 
& &   \\ \hline
\multirow{2}{*}{Euclidean lattice~\cite{Eisert_2010_review,Laflorencie_2016_review}}& Gapped,\, DOS$=0$ & $\sim L_{A}^{d-1}$ \\
& Gapless,\, DOS$>0$ & $\sim L_{A}^{d-1}\log L_{A}$ \\ \hline
\multirow{2}{*}{Fractal lattice~\cite{zhou2024quantum}}& Gapped,\, DOS$=0$ & $\sim L_{A}^{d_{\rm bf}}$\\
& Gapless,\, DOS$>0$ & $\sim L^{d-1}_{A} \log L_{A}$\\  \hline
\multirow{2}{*}{ hyperbolic lattice} & Gapped,\, DOS$=0$ & $\sim L_{A}$  \\
& Gapless,\, DOS$>0$ & $\sim L_{A}$  \\ 
\end{tabular}
\end{ruledtabular}
\end{table}

To achieve our research objectives, our methodology begins with the application of the vertex inflation method~\cite{Boyle_2020_C,Jahn_2020_C,Chen_2023_S}. This method is instrumental in creating a hyperbolic lattice configuration on the Poincar\'{e} disk, which serves as the foundational structure for our computational study. 
Following the lattice creation, we employ the Haydock recursion method~\cite{Mosseri_2023_D,Haydock1972,Haydock1973,Haydock1975} to compute DOS within this hyperbolic framework. This computational technique is well-suited for handling the complex geometries inherent in hyperbolic lattices, providing a detailed characterization of electronic states and their distribution~\cite{Mosseri_2023_D}. 
Subsequently, we proceed to obtain the eigen spectrum of non-sparse reduced density matrices via exact diagonalization and various kinds of bi-partitions between the two subsystems. To obtain the scaling behaviors, we perform  finite-size scaling analyses, which  enables us to extrapolate our findings across different subsystem sizes, revealing how entanglement quantities scale with the boundary of the subsystem.  
Furthermore, a central aspect of our investigation involves exploring correlations between the coefficient of the area law, bulk gap, and DOS. 
As  hyperbolic lattice can be experimentally realized through various techniques, it will be interesting to experimentally measure entanglement on hyperbolic lattices via, e.g., phononic platform~\cite{Lin2024}.
Interestingly, the area law of both gapless and gapped systems implies  that the matrix product states (MPS) and projected entangled-pair states (PEPS)~\cite{ORUS_2014,Cirac_2021_M,Xiang_2023}  may be potentially efficient in simulating quantum spin liquids with gapless spinons with finite DOS on hyperbolic lattice.

This paper is arranged as follows: In Sec.~\ref{sec:fundamentals}, we specify the construction of hyperbolic lattices and provide a brief summary of studying free-fermion entanglement entropy. Next in Sec.~\ref{sec:gapless} we study EE of gapless free-fermion systems with finite DOS and the dependence of scaling coefficient on DOS while in Sec.~\ref{sec:gapped}, we study EE of gapped free fermions on hyperbolic lattices. Finally, we summarize our findings in Sec.~\ref{sec:conclusion} and discuss their potential applications. Additionally, we detail the hyperbolic lattice setup and discuss the volume law in Appendix~\ref{app:lattice}, provide supplemental data of EE in Appendix~\ref{app:fit} and review the approach to compute DOS in Appendix~\ref{app:DOS}. We provide the finite-size scaling analysis in Appendix~\ref{app:finitesizescaling} and the analysis of super-area law in Appendix~\ref{app:superarea}. We discuss the asymptotic behavior of the coefficient of the area law in Appendix~\ref{app:asymtotic}.

\section{Preliminaries}
\label{sec:fundamentals}

\subsection{Tessellations of plane}
\begin{figure}[b]
	\includegraphics[width=\columnwidth]{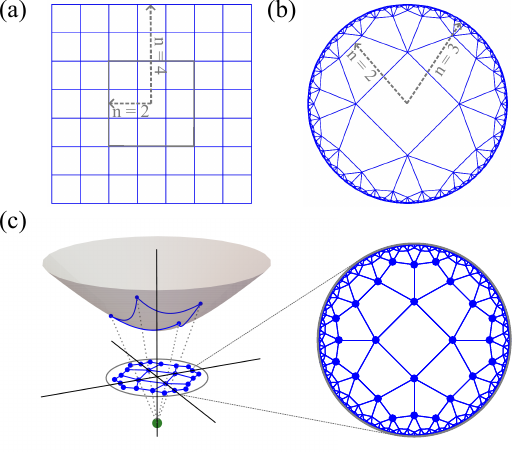}
	\caption{\label{fig:tilings} Tessellations of a two-dimensional plane and projection of hyperbolic lattice. (a) Euclidean $\{4,4\}$ lattice. (b) Hyperbolic $\{4,6\}$ lattice. The gray labeled dashed line denotes the order $n$ of the ring. (c) Projection of $\{4,6\}$ lattice onto a Poincar\'{e} disk. A site on the hyperboloid $\hat{z}^2-\hat{x}^2-\hat{y}^2=1$ is projected onto a unit disk on the $\hat{z}=0$ plane by intersecting it with a line drawn through $(0,0,-1)$.}
\end{figure}

In the beginning, we introduce the tessellations (or tilings) of the Euclidean and hyperbolic plane. 
A two-dimensional plane can be tessellated by regular polygons, denoted by the Schl\"afli symbol $\{p,q\}$~\cite{Boettcher_2022_C}, where the integers $p$ and $q$ represent that the plane is tessellated by regular $p$-edges polygons, with each lattice site having coordination number $q$.
For instance, as demonstrated in Fig.~\ref{fig:tilings}(a), each square has edges $p=4$ and each lattice site has coordination number $q=4$ for square lattice $\{4,4\}$. 
For the two-dimensional plane with Euclidean geometry, $p,q$ should satisfy the constraint $(p-2)(q-2)=4$, which means that there are only three possible tessellations, including the triangular lattice $\{3,6\}$, the square lattice $\{4,4\}$, and the hexagonal lattice $\{6,3\}$. In addition, when $p$ and $q$ satisfy $(p-2)(q-2)>4$, these tessellations can be adopted to discretize the hyperbolic plane and Fig.~\ref{fig:tilings}(b) demonstrates $\{4,6\}$ lattice.

Before constructing hyperbolic lattices, We need to specify the coordinates under which we are handling our studies. To assign a complex coordinate to each lattice site, we employ a conformal disk model of hyperbolic space, i.e., Poincar\'{e} disk as shown in the right-hand side of Fig.~\ref{fig:tilings}(c). 
By using this conformal map, the lattice is embedded in a unit disk $\mathbb{D}=\{z\in \mathbb{C}, \lvert z\rvert < 1\}$ with metric 
\begin{align}
	\label{metric}
	ds^2=(2\kappa)^2\frac{\lvert dz\rvert ^2}{(1-\lvert z\rvert^2)^2}\,,
\end{align}
where $\kappa$ is the constant radius curvature and its corresponding constant curvature is $K = -\kappa^{-2}$. 
From Eq.~(\ref{metric}), the geodesic distance $\sigma$ between two sites $z$ and $z^\prime$ on the Poincar\'{e} disk is given by 
\begin{align}
	\sigma(z,z^\prime)=\kappa\ \text{arcosh}\left(1+\frac{2\lvert z - z^\prime\rvert^2}{(1-\lvert z\rvert^2)(1-\lvert z^\prime \rvert^2)}\right)\,,
\end{align}
where $z$ denotes a site on the disk with complex coordinate $z=x+\mathrm{i} y=re^{\mathrm{i}\phi}$.

\subsection{Hyperbolic lattice construction and the exponential wall}
\label{subsec:lattice}
\begin{figure}[b]
	\includegraphics[width=1\columnwidth]{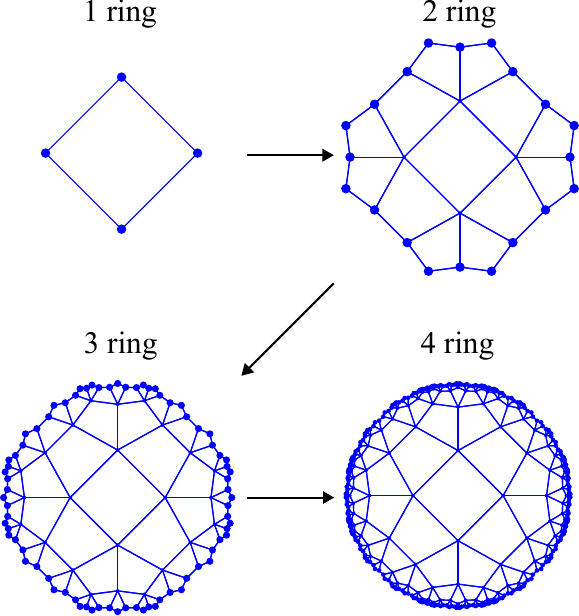}
	\caption{\label{fig:lattice} Generating procedure of $\{4,5\}$ lattice with 1,2,3,4 rings using vertex inflation method. The bold sites of each lattice highlight the iteratively attached outermost ring, i.e., the boundary of that lattice.}
\end{figure}

Next, we consider using the regular tilings to generate hyperbolic lattices. By adopting the vertex inflation method (or vertex-inflation tiling procedure)~\cite{Boyle_2020_C,Jahn_2020_C,Chen_2023_S}, we can effectively generate Euclidean and hyperbolic lattices of various rings where the sites are located. To obtain a finite $\{p,q\}$ lattice, we initially generate a regular $p$-edges polygon at the center of the Poincar\'{e} disk, labeled as the first ring and then attach new rings to it iteratively. In Fig.~\ref{fig:lattice} we show the generating procedure of $\{4,5\}$ lattice, where the bold sites denote the outermost ring that generated in each iterative step.
By repeating this process, we can successively enlarge the size of the lattice based on the outermost ring, allowing us to obtain an arbitrarily large lattice with any number of rings. More detailed information on this procedure can be found in Appendix~\ref{app:lattice}. 

In the following, we use $\{p,q,n\}$ rather than $\{p,q\}$ to label a concrete finite hyperbolic lattice, i.e., flake, for numerical computations, where the integer $n$ represents the number of rings included in the lattice, as shown in Fig.~\ref{fig:tilings}(a) and (b) plotted by the dash line. 
An important feature of hyperbolic lattice is that the total number of lattice sites $N$ increases exponentially with the number of ring $n$ as $N \sim \lambda^n $, where $\lambda$ is a parameter depending on specific $\{p,q\}$. In contrast, for Euclidean lattices, $N\sim n^2$. 
Additionally, the number of sites $N_{boun}$ on the outermost ring of the hyperbolic lattice, which corresponds to the boundary, also increases exponentially with $n$ for large $n$, whereas in Euclidean lattices, it increases linearly as $N_{boun} \sim n$ . 
A brief proof of these properties can be found in Appendix~\ref{app:lattice}, highlighting the fundamental differences between the two geometry. These properties all bring difficulties for numerical computations. 

\subsection{Partition of subsystems on the hyperbolic lattice}
\begin{figure}[b]
	\includegraphics[width=1\columnwidth]{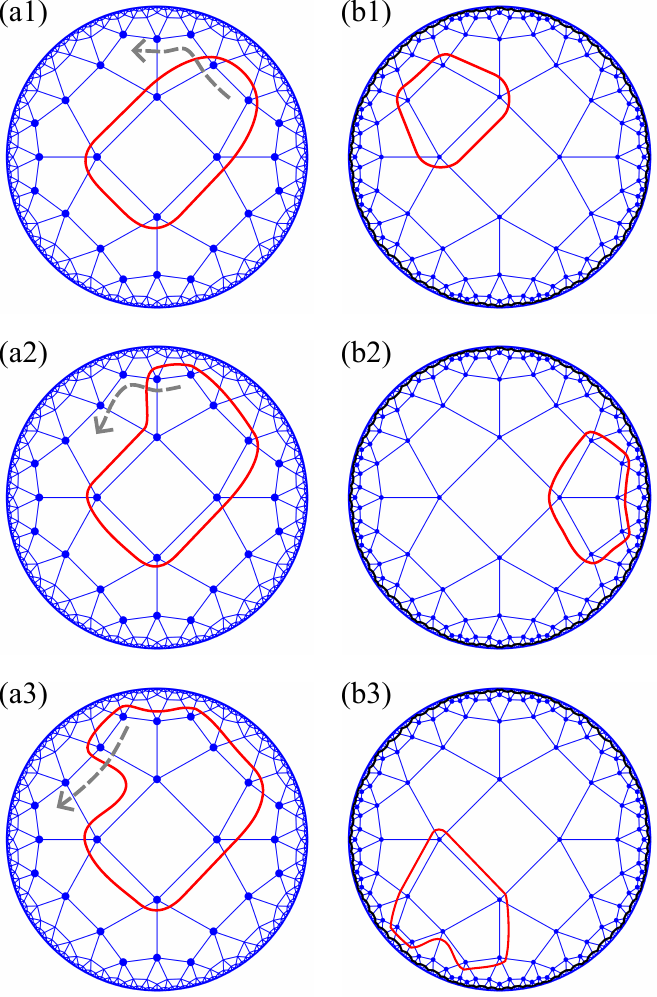}
	\caption{\label{fig:partition} Partition of subsystems on $\{4,5,6\}$ lattice. (a) Subsystems generated through \textit{partition} $\text{\romannumeral 1}$ that is adopted in the main text. (b) Subsystems generated through \textit{partition} $\text{\romannumeral 2}$. Here we generate random subsystems of specific size 4 (b1), 6 (b2) and 8 (b3) within the region denoted by the the black line. The number of bonds connecting sites inside the subsystem to sites outside, which are cut by the red line, are defined as the boundary $L_A$ of the subsystem.}
\end{figure}
Since the choice of subsystem affects EE, we now turn to specify our partition methods. When partitioning subsystems to study EE, we need to choose the largest possible subsystems while keeping them as far from the boundary as possible to minimize finite-size effect. 
However, as explained in Sec.~\ref{subsec:lattice}, $N$ and $N_{boun}$ grow exponentially with $n$, making it difficult to have a relatively large bulk.
We define $R_i$ as the shortest discrete graph path from a bulk site $i$ to the boundary. Sites with $R$ larger than a certain threshold $R_{min}$ can be chosen to form a single-connected region as $A$, thereby positioning the subsystem on the inner rings of the lattice.
Regarding the symmetry of the subsystems, on Euclidean lattices, subsystems are typically chosen as a series of polygons similar to the overall system.
However, the symmetry of hyperbolic lattice, described by the \textit{triangle} group and the \textit{Fuchsian} group, is non-Abelian~\cite{Maciejko_2021_HBT,Maciejko_2022_A,Lenggenhager_2023_N}. Consequently, the subsystems cannot maintain the same symmetries as on the Euclidean lattice.

Therefore, we employ two different partition methods in this work. We first generate a lattice of fixed size, within which we choose the sites of the innermost ring as the initial subsystem $A$, and then successively increase its size by adding sites of the adjacent ring to it in a clockwise or anti-clockwise direction. This iterative procedure, which generates a sequential series of subsystems, is visualized in Fig.~\ref{fig:partition}(a), and is referred to as \textit{partition} $\text{\romannumeral 1}$.
Additionally, we also conduct a random partition of subsystem. We determine a minimum $R_{min}$ for a considered lattice $\{p,q,n\}$ and generate subsystem within this region. We first randomly choose a $p$-edges polygon, then enlarge it by successively adhering $p$-edges polygons around sites on the boundary of the subsystem to it and repeat this procedure until it reaches a specific size. This partition method is referred to as \textit{partition} $\text{\romannumeral 2}$ and can be visualized in Fig.~\ref{fig:partition}(b). 
Since the partitions do not consistently preserve the symmetries of the subsystems, we find that through \textit{partition} $\text{\romannumeral 2}$ the symmetries do not significantly affect the numerical results of EE in practical computations. 
 {In the remaining part of the main text, we consistently exhibit the results of EE computed through \textit{partition} $\text{\romannumeral 1}$ on some lattices and provide the supplemental data in Appendix~\ref{app:fit} for more details of both  \textit{partition} $\text{\romannumeral 1}$ and  \textit{partition} $\text{\romannumeral 2}$.}

\subsection{Entanglement entropy  and Widom conjecture}
Next, we concisely review some basic algebras for computing the entanglement of free-fermion systems. A useful relevant material can be found in the supplementary note of Ref.~\cite{Lin2024}. For a many-body system with ground state $\left| GS \right> $, its density matrix is $\rho = \left| GS \right> \left< GS \right|$. We partition the system into two parts as subsystem $A$ of the overall system and its complementary $B$ in real space, and obtain reduced density matrix $\rho_A$ of subsystem $A$ by tracing over $B$: 
\begin{align}
	\rho_A = \mathrm{Tr}_B \left| GS \right> \left< GS \right| = \frac{1}{\mathcal{Z}}\mathrm{exp}(-H^E)\,,
\end{align}
where $\mathcal{Z}$ is a normalization constant and $H^E$ is the entanglement Hamiltonian, from which we can obtain EE~\cite{Haldane_2008_E,Fidkowski_2010_E,Lee_2014}. If we consider free-fermion systems, $H^E$ has quadratic form~\cite{Klich_2006,Ingo_Peschel_2003,Lee_2015} $H^E=\sum_{i,j\in A}{c_{i}^{\dagger}h_{ij}^{E}c_j}$, 
where $c^{\dagger}_i$ and $c_i$ represent the fermionic creation and annihilation operators at site $i$ respectively. 
Additionally, we can rewrite EE as a trace of matrix-function. Consider the correlation matrix $C^A_{ij}=\left< GS \right| c_{i}^{\dagger}c_j \left| GS \right>$ of subsystem $A$ which can be obtained by projection operators $C^A=\hat{R}\hat{P}\hat{R}$ where $\hat{R}=\sum_{i\in A}{\left| i \right> \left< i \right|}$ and $\hat{P}=\sum_{k \in occ.}{\left| k \right> \left< k \right|}$, the EE can be calculated by~\cite{Ingo_Peschel_2003,Lee_2014,Lee_2015,Lai_2015_E,Crampé_2019,Chen_2021_nonhermi,Lee_2022}:
\begin{align}
	\label{ee}
	S_A=\mathrm{Tr}_Af(C^A)=\mathrm{Tr}_Af(\hat{R}\hat{P}\hat{R})\,,
\end{align}
where $f(t)=-t\log t-(1-t)\log(1-t)$. 
Hence we obtain EE of subsystem $A$. 

Meanwhile, for gapless systems with codimension-1 Fermi surface, the Widom conjecture  provides an analytical result of EE~\cite{Gioev_2006,Leschke_2014}:
\begin{align}
	\label{widom_conjecture}
	S_A= \frac{L_{A}^{d-1}\log L_A}{(2\pi )^{d-1}}\frac{1}{12}\iint_{\partial \Gamma \times \partial \Omega}{\left| \bm{n_r}\cdot \bm{n_p} \right|dS_rdS_p}\,,
\end{align}
where $\partial\Gamma$ and $\partial \Omega$ denote the boundaries of the Fermi sea and the subsystem we consider, $\bm{n_p}$ and $\bm{n_r}$ denote the exterior unit normals of these boundaries. Since the presence of codimension-1 Fermi surface implies finite DOS of the system, Eq.~(\ref{widom_conjecture}) also relates the DOS to the scaling coefficient. If the codimension of the Fermi surface is higher than one, the leading term of EE exhibits area law scaling behavior, as seen in the Dirac point of tight-binding model on the honeycomb lattice~\cite{Barthel_2006_E,Levine_2008_Area,Ding_2008_Subarea}.
However, the validity of Eq.~(\ref{widom_conjecture}) requires a Euclidean metric with Abelian translation symmetry and thus is not naturally applicable in the hyperbolic geometry, so we aim to provide numerical evidence in this paper.

\section{Entanglement entropy scaling of gapless free-fermion systems with finite DOS}
\label{sec:gapless}
\begin{figure*}
	\includegraphics[width=1\linewidth]{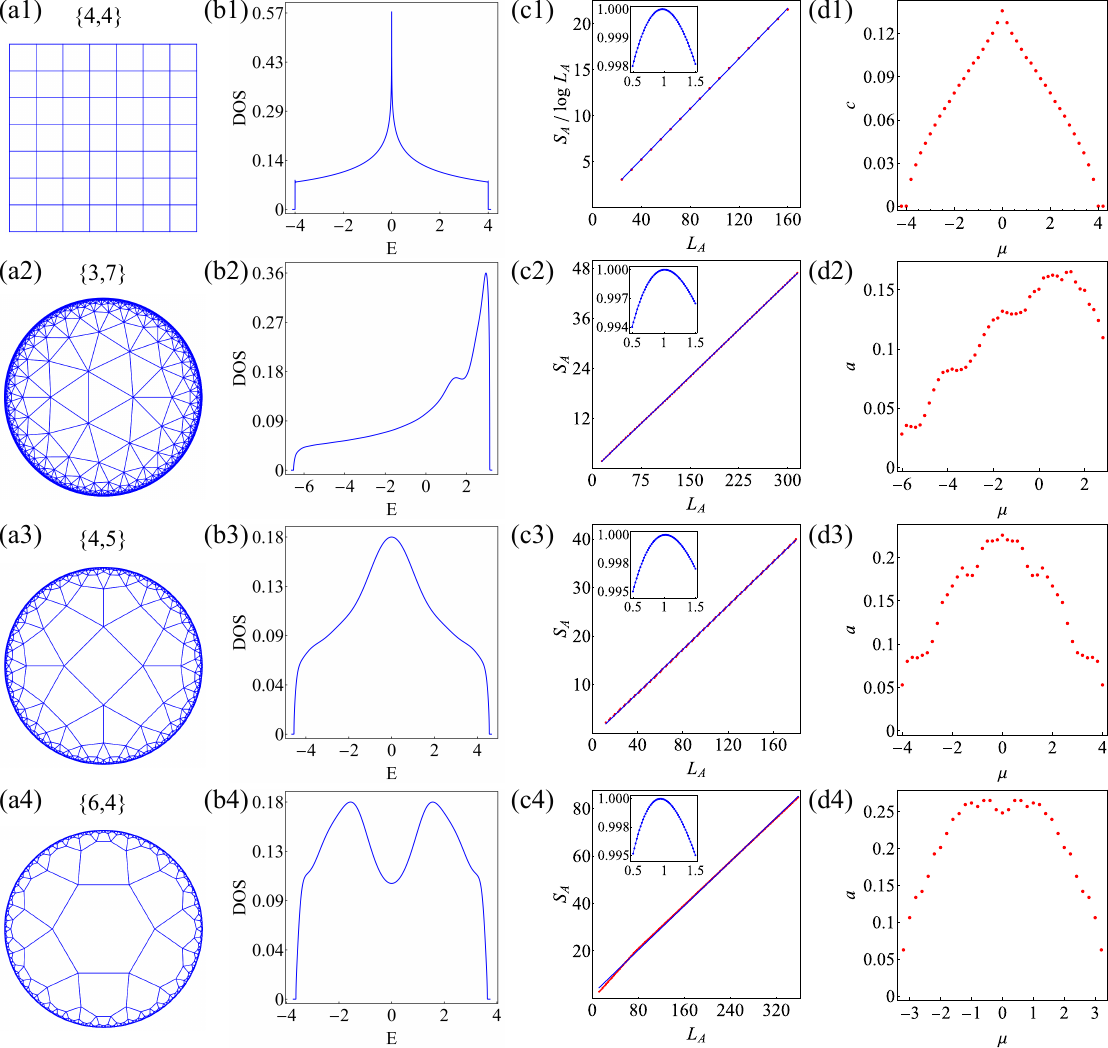}
	\caption{\label{fig:gapless_entropy} Linear fit of EE and dependence of scaling coefficients on DOS for Euclidean $\{4,4\}$ and hyperbolic $\{3,7\}$, $\{4,5\}$, $\{6,4\}$ lattices (Row 1 to 4, respectively) of gapless systems with Hamiltonian $H_1$. Column (a) shows the lattices. Column (b) shows DOS computed by Haydock recursion method which gives the DOS in the thermodynamical limit of Hamiltonian $H_1$, with details in Appendix~\ref{app:DOS}. Column (c) shows the linear fit of EE and boundary (\textit{partition} $\text{\romannumeral 1}$ is taken and results of \textit{partition} $\text{\romannumeral 2}$ are exhibited in Appendix~\ref{app:fit}). The insets show the coefficient of determination $R^2$ as a function of $\alpha$. In column (b) and (c) we set $t=1$ and $\mu=0$. Column (d) shows the dependence of coefficients $a$ and $c$ on DOS. Column (c) and (d) are numerically computed on $\{4,4,40\}$ (6400 sites), $\{3,7,9\}$ (17328 sites), $\{4,5,6\}$ (5400 sites) and $\{6,4,5\}$ (10086 sites) lattices respectively.}
\end{figure*}

\subsection{Numerical study of DOS}
In this section, we numerically study the scaling behavior of EE of gapless free-fermion systems with finite DOS on hyperbolic lattice. To begin with, we consider the gapless systems with a one-orbital tight-binding model:
\begin{align}
	\label{gapless hamiltonian}
	H_1=-t\sum_{\left< ij \right>}{\left( c_{i}^{\dagger}c_j+h.c. \right)}-\mu \sum_i{c_{i}^{\dagger}c_i},
\end{align}
where $\left<ij\right>$ denotes the nearest-neighboring sites, $t$ is the hopping amplitude and $\mu$ is the chemical potential. 
First, we should verify that the Hamiltonian $H_1$ is indeed gapless. 
We notice that the DOS obtained through exact diagonalization for $10^4$ sites still exhibits finite-size effect, and thus it's insufficient to verify whether the system is gapless or not through it. 
Consequently, we analyze DOS in the thermodynamical limit through the Haydock recursion method~\cite{Mosseri_2023_D,Haydock1972,Haydock1973,Haydock1975}.

One can calculate local DOS $\rho_i(E)$ at a site $i$ through Green's function:
\begin{align}
	\label{ldos}
	\rho_i(E) = -\lim_{\epsilon\rightarrow 0^+} \frac{1}{\pi}\mathrm{Im}\left< i \lvert G(E+\mathrm{i}\epsilon) \rvert i \right>\,,
\end{align}
where $\left| i \right>$ is the state we consider and the Green's function is $G(E)=1/ (E-H)$. The diagonal element of $G$ can be expanded in continued-fraction:
\begin{align}
    G_{ii}\left( E \right) =\frac{1}{E-a_1-\frac{b_{1}^{2}}{E-a_2-\frac{b_{2}^{2}}{\cdots}}}\,,
\end{align}
where the rational coefficients $a_n$ and $b_n$ can be numerically computed by the underlying Hamiltonian matrix $H$ through specific recursive relation. After introducing a proper fraction termination, we obtain $\rho_i(E)$ which is also DOS for regular tilings up to a normalization factor~\cite{Mosseri_2023_D}. 
By using this method, we confirm that the Hamiltonian $H_1$ is indeed gapless on lattices that we study here, as shown in Fig.~\ref{fig:gapless_entropy}(b1-b4). One can refer to Appendix~\ref{app:DOS} for more details of this method and numerical results.

\subsection{Numerical evidence of area law scaling behavior of EE}
\label{subsec:gapless_area_law}
To proceed further, we use our approaches detailed in Sec.~\ref{sec:fundamentals} to compute EE on various lattices, including both Euclidean and hyperbolic. In Fig.~\ref{fig:gapless_entropy}(c2-c4) we show the results computed on $\{3,7\}$, $\{4,5\}$, $\{6,4\}$ lattices. Additionally, in Fig.~\ref{fig:gapless_entropy}(c1), we also include EE computed on Euclidean $\{4,4\}$ lattice for comparison. 
More numerical results through different partition methods on different lattices are detailed in Appendix~\ref{app:fit}. The finite-size scaling analysis can be found in Appendix~\ref{app:finitesizescaling}. The numerical analysis of super-area law is presented in Appendix~\ref{app:superarea}.

First, in the Euclidean case, the EE of gapless systems with finite DOS exhibits super-area law, corresponding to our results computed on $\{4,4\}$ lattice in Fig.~\ref{fig:gapless_entropy}(c1), where we anticipate the scaling function $S_A / \log L_A=cL_A^\alpha+d$.
When we turn to the hyperbolic case, our most surprising finding is that the EE of gapless systems with finite DOS is proportional to the length of the boundary of subsystem $A$. We anticipate that the scaling of EE should have $S_A=a L_A^\alpha + b$. By using the coefficient of determination $R^2$, we find that $\alpha\approx 1$ is the optimal fit closest to $1$, as shown in Fig.~\ref{fig:gapless_entropy}(c2-c4). The blue lines in Fig.~\ref{fig:gapless_entropy}(c) show the fitting functions with $\alpha = 1$. This result indicates that the EE of gapless systems with finite DOS on hyperbolic lattices satisfies the area law by definition: 
\begin{align}
	\label{gapless_area_law}
	S_A = a L_A +\cdots\,,
\end{align}
where $L_A$ represents the total number of bonds connecting a site inside the subsystem to a site outside the subsystem which are cut by the boundary of the subsystem $A$, which is consistent with the Euclidean case and visualized in Fig.~\ref{fig:partition}.  
As the number of sites on the boundary can grow linearly with the number of sites in the subsystem in the thermodynamical limit, we discuss the volume law of EE in Appendix~\ref{app:lattice}.

With more numerical computations, as illustrated in Appendix~\ref{app:fit}, and through the numerical analysis that excludes of the possibility of super-area law scaling presented in Appendix~\ref{app:superarea}, we further confirm the existence of area law of EE for gapless ground states with finite DOS on hyperbolic lattice. 
Recent Reference~\cite{chen_2023_adscft} experimentally simulates weakly-coupled scalar field to study AdS/CFT correspondence on hyperbolic lattice. In this work, the EE behavior for entanglement-wedge subsystems of the bulk scalar field satisfies the Ryu-Takayanagi (RT) formula~\cite{RT} for the connection between the EE of boundary CFT and geometry of the hyperbolic lattice, a result that also has physical understanding~\cite{RT,witten,Klebanov1998}. 
Furthermore, we want to ask why this exotic area law Eq.~(\ref{gapless_area_law}) of gapless free-fermion systems with finite DOS appears in hyperbolic case. 
The analytical formula of the EE is based on the Widom conjecture of the asymptotic behaviors of Toeplitz matrices on the Euclidean lattice. Due to the absence of Euclidean translation invariance on hyperbolic lattice, the future analytical understanding of EE could be associated with studying conjecture of correlation matrices with symmetry of hyperbolic lattice.

Moreover, following Swingle's \textit{mode-counting} argument~\cite{Swingle_2010_E}, for free-fermion systems with codimension-1 Fermi surface, EE can be obtained by counting the contributions of $1$d fermionic gapless modes near the Fermi surface perpendicular to the boundary of the subsystem in real space, where each fermionic gapless mode contributes $\log L_{A}$ to EE by adopting the calculation of CFT. Then, we obtain that EE satisfies $S_{A}\sim L_{A}^{d-1}\log L_{A}$ in Euclidean case.
On hyperbolic lattice, the Swingle's \textit{mode-counting} picture is invalid due to the absence of ``Fermi surface'' of the usual definition. 
If we can stack and count the contribution of the infinite fermionic gapless modes near the generalized ``Fermi surface'' for EE, we can obtain the scaling behavior of EE for hyperbolic systems. 
However,  there is a lack of a realizable stacking and counting way on hyperbolic lattice. 
According to Eq.~(\ref{ee}), EE depends on the projectors $\hat{P}$ and $\hat{R}$. HBT provides an insight for us into the parameterization of the generalized hyperbolic momentum space and non-Abelian Bloch states~\cite{Maciejko_2021_HBT,Maciejko_2022_A,Lenggenhager_2023_N}. Therefore, our numerical simulation raises questions and challenges for HBT to obtain a generalized Widom conjecture and Swingle's \textit{mode-counting} picture for hyperbolic lattices, as well as the expressions of $\hat{P}$ and $\hat{R}$ from the parameterized momentum space.

The scaling behavior of EE is related to the non-local properties of the systems. Due to the absence of the logarithmic correction of EE in Eq.~(\ref{gapless_area_law}), we realize that the gapless fermions on hyperbolic lattices should have exotic behavior owing to its non-trivial underlying hyperbolic geometry, and the study of this area law may provide perspective of entanglement for HBT as discussed above. 
Additionally, as hyperbolic geometry suppresses entanglement,  it is worth investigating the asymptotic behavior of EE with respect to $q$ and we discuss this in Appendix~\ref{app:asymtotic}. 
In the forthcoming Sec.~\ref{subsection_scaling_coef}, we will continue to discuss our numerical findings, especially focusing on the scaling coefficient.

\begin{figure}[b]
	\includegraphics[width=0.9\columnwidth]{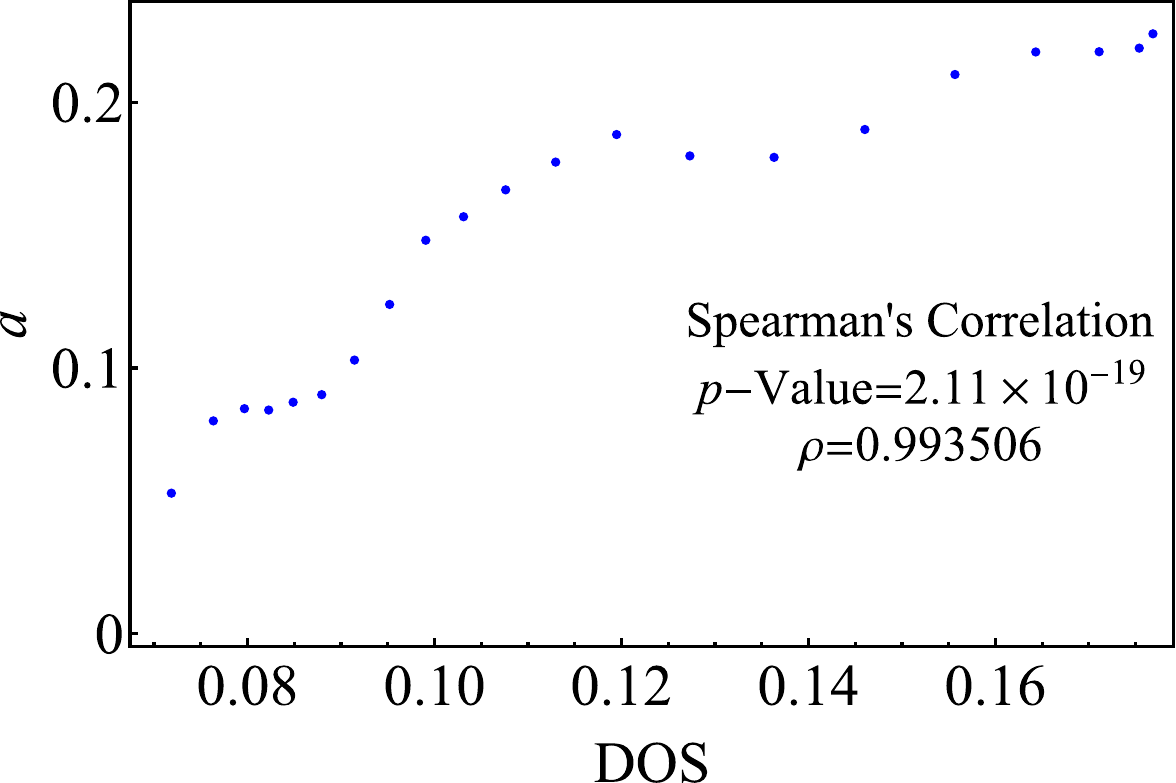}
	\caption{\label{fig:EE_DOS_cor} Dependence of coefficient $a$ on DOS of $\{4,5,6\}$ lattice, computed for Hamiltonian $H_1$ with $t=1$ and $\mu\in [0,4]$. The $p$-value approaching $0$ and $\rho$ approaching $1$ of the Spearman's correlation verifies the approximately positive correlation between scaling coefficient and DOS, suggesting that a generalized Widom conjecture may exist.}
\end{figure}

\subsection{Numerical study of scaling coefficient and possibility of a generalized Widom conjecture}\label{subsection_scaling_coef}
In the Euclidean case, we know from Eq.~(\ref{widom_conjecture}) that the scaling coefficient of the super-area law is analytically determined by the flux factor $ \left| \bm{n_r}\cdot \bm{n_p} \right|$ reflecting geometry of the codimension-1 Fermi surface and thus the scaling coefficient changes according to DOS, as visualized in Fig.~\ref{fig:gapless_entropy}(d1). 
Because hyperbolic lattice allows for generalized momentum space, we question whether the DOS can influence the coefficient $a$ in area law Eq.~(\ref{gapless_area_law}) following the Euclidean scenario. 

We compute EE for Hamiltonian $H_{1}$ with $t=1$ and different chemical potential $\mu$ on different hyperbolic lattices. Fig.~\ref{fig:gapless_entropy}(d2-d4) shows the dependence of the scaling coefficient $a$ on $\mu$. 
Compared to the DOS computed in Fig.~\ref{fig:gapless_entropy}(b2-b4), we can directly see that the scaling coefficient $a$ is correlated to the DOS. 
In Fig.~\ref{fig:EE_DOS_cor}, we present the result of $a$ computed on $\{4,5,6\}$ lattice as an example, where the Spearman's correlation indicates the approximately positive correlation between $a$ and the DOS. 
Notably, from Fig.~\ref{fig:gapless_entropy}, we can see that $a$ and the DOS do not completely coincide. This discrepancy might also be due to the finite-size effect, as the EE computed here is obtained from a finite lattice while the DOS is obtained in the thermodynamical limit.

Remarkably, the scaling coefficient $a$ is non-universal and influenced by many factors such as partition and lattice configuration, but the approximately positive correlation between $a$ and the DOS implies that a generalized momentum space and Fermi surface might play a role in determining $a$, similar to its Euclidean counterpart.
In the Euclidean system, the validity of the Widom conjecture and Swingle's \textit{mode-counting} picture need a Euclidean metric and the momentum space with dimension equal to real-space dimension due to the flux factor $\left| \bm{n_r}\cdot \bm{n_p} \right|$ in Eq.~(\ref{widom_conjecture}) counting the number of fermionic modes perpendicular to the real space boundary of the subsystem. 
In fact, the translation group of hyperbolic lattice is typically non-Abelian, resulting in the existence of higher-dimensional ($d\ge2$) irreducible representations of translation group and non-Abelian Bloch states. Meanwhile, even for the $1$d irreducible representations, the dimension of the generalized momentum space can be $ d>2$~\cite{Maciejko_2021_HBT,Maciejko_2022_A,Lenggenhager_2023_N}, which is larger than the spatial dimension of the lattice, thus Swingle's argument breaks down directly. To exactly obtain a description of reciprocal space of hyperbolic lattice, one needs to know about the higher-dimensional representations. It is an open question that whether we can obtain a generalized Widom conjecture and Swingle's \textit{mode-counting} picture for hyperbolic lattice.

\begin{figure*}
	\includegraphics[width=1\linewidth]{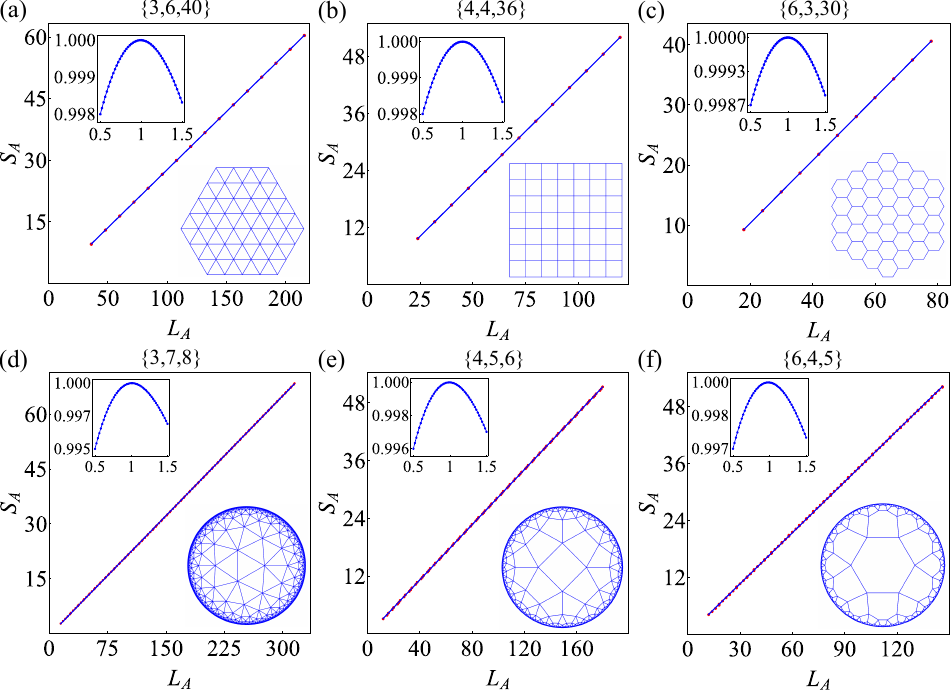}
	\caption{\label{fig:gapped_entropy_fit} Linear fit of EE in the gapped case with Hamiltonian $H_2$ on $\{3,6,40\}$ (4800 sites) (a), $\{4,4,36\}$ (5184 sites) (b), $\{6,3,30\}$ (5400 sites) (c), $\{3,7,8\}$ (6615 sites) (d), $\{4,5,6\}$ (5400 sites) (e) and $\{6,4,5\}$ (10086 sites) (f) lattices which have two orbitals at a site. The fittings show the area law of EE. The insets show $R^2$ as a function of $\alpha$. Such a linear dependence of $S_A$ on $L_A$ is consistent with the Euclidean case. The hopping amplitude of $H_2$ are set to $t_1=1$ and $t_2=1$.}
\end{figure*}

\section{Entanglement entropy scaling of gapped free-fermion systems}
\label{sec:gapped}
\begin{figure}[b]
	\includegraphics[width=\columnwidth]{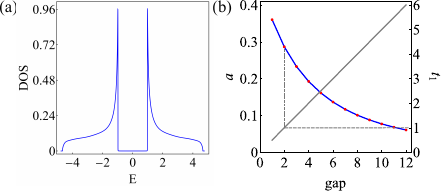}
	\caption{\label{fig:45_entropy_dos_gapped}DOS and EE computed on $\{4,5\}$ lattice for Hamiltonian $H_2$. (a) DOS computed by Haydock recursion method. The hopping amplitude $t_1=1$ and $t_2=1$ lead to gapped region $[-1,1]$. (b) Scaling coefficient $a$ of area law varies with different energy gaps, computed on $\{4,5,6\}$ lattice. The gray line shows the energy gap modulated by $t_1$.}.
\end{figure}

In this section, we study EE in gapped systems. We consider the gapped systems by studying a two-orbital tight-binding model:
\begin{align}
	\label{gapped hamiltonian}
	H_2=-\sum_i{t_1\left( c_{s,i}^{\dagger}c_{p,i}+h.c.\right)}-\sum_{\left<ij\right>}{t_2\left( c_{s,i}^{\dagger}c_{s,j}-c_{p,i}^{\dagger}c_{p,j} \right)}\,,
\end{align}
where $c^\dagger_{s(p),i(j)}$ represents fermionic creation operator at the $s(p)$-orbital of site $i(j)$. $t_1$ and $t_2$ are hopping amplitudes. 
We can still use Haydock recursion method to compute DOS and verify that $H_2$ is gapped as we did in Sec.~\ref{sec:gapless}. 
For instance,in Fig.~\ref{fig:45_entropy_dos_gapped}(a), we show the DOS of Hamiltonian $H_2$ with $t_1=1$ and $t_2=1$ on $\{4,5\}$ lattice, which lead to a gapped region $[-1,1]$.

Next, we turn to study EE in gapped case. On Euclidean lattice, EE of gapped systems scales as area law $S_A=a L_A^\alpha + \cdots$. 
As an analogy, we also use the fitting function $S_A=aL_A^\alpha+b$ for the case on hyperbolic lattice. 
In Fig.~\ref{fig:gapped_entropy_fit}, we show results of EE computed on both Euclidean and hyperbolic lattices. The chosen hyperbolic lattices $\{3,7\}$, $\{4,5\}$ and $\{6,4\}$ have one more adjacent site per lattice site compared to their Euclidean counterparts $\{3,6\}$, $\{4,4\}$ and $\{6,3\}$ respectively. 
The numerical results consistently show that when the the system is gapped, the optimal fit is obtained with $\alpha\approx 1$ where $R^2$ is closest to $1$. The blue lines in Fig.~\ref{fig:gapped_entropy_fit} show the fitting functions with $\alpha = 1$. This means that the EE scales linearly with the subsystem's boundary $L_A$:
\begin{align}
	\label{gapped_area_law}
	S_A = a L_A + \cdots\,.
\end{align}
Therefore, EE still scales according to area law in gapped systems on hyperbolic lattice.

Additionally, on Euclidean lattices, the coefficient $a$ decreases as the energy gap increases. This leads us to question whether the energy gap is related to the behavior of EE. In Fig.~\ref{fig:45_entropy_dos_gapped}(b), we study the relation between EE and energy gap on $\{4,5,6\}$ lattice. We modulate $t_1$ and thus change energy gap of $H_2$ from $1$ to $12$ and compute EE. We find that $a$ is negatively correlated with the system's energy gap. 
Analytical work on the one-dimensional gapped system has provided a rigorous relationship between the coefficient $a$ and the energy gap~\cite{Hastings_2007, Eisert_2010_review}. However, the exact relationship between the scaling coefficient $a$ and the energy gap is still a difficult question in dimension $d\ge2$. 
In our results, we do not find a functional relationship that can physically explain the relationship between the coefficient $a$ and the energy gap in the hyperbolic case, but the observed negative monotonic relationship between them suggests a similarity to the Euclidean case.

Overall, our numerical data computed in gapped systems demonstrates that the EE scales according to area law as in Eq.~(\ref{gapped_area_law}). This aligns with our expectations from the Euclidean case, suggesting that the gapped scenario in the hyperbolic case is not particularly unique.

\section{Discussions}
\label{sec:conclusion}
In this paper, we have numerically studied the scaling behavior of entanglement entropy of gapped free fermions as well as gapless free fermions with finite DOS on hyperbolic lattice. 
We find that for both gapped and gapless systems, the EE scales according to a rigorous area law scaling $S_A = a L_A + \cdots$. 
Although the gapped case fulfills our expectation in Euclidean geometry, the super-area law in gapless systems breaks down in contrast.
Additionally, the scaling coefficient of area law in gapless systems is positively correlated to the DOS. 
This scaling behavior of EE is unique in hyperbolic geometry. 
On Euclidean lattice, the super-area law of gapless free fermions with finite DOS demonstrates that the entanglement is enhanced by the fermionic statistics and the quantum correlation of the infinite fermion modes near the Fermi surface~\cite{Gioev_2006,Swingle_2010_E}. 
These observations show that the underlying geometry can significantly influence the entanglement behavior of ground states of free-fermion systems, similar to our previous findings in fractal geometry~\cite{zhou2024quantum}, as summarized in Table.~\ref{tab:eegeometry}. 
Compared to Euclidean case, the area law reveals the presence of exotic properties of fermions on hyperbolic lattice.  
The further study of this area law might raise the question for a generalized conjecture of correlation matrices with symmetry of hyperbolic lattice, while the possibility of generalization of Swingle's argument for EE to the hyperbolic case through HBT also merits future research~\cite{Maciejko_2022_A,Maciejko_2021_HBT,Lenggenhager_2023_N}.

Notably, many studies suggest the relationship between the entanglement and the geometry of AdS space, particularly in the context of hyperbolic lattice~\cite{chen_2023_adscft,Gendiar_2020_area,Nishino_2024_holographic,2024arXiv240917235S}.   
Ref.~\cite{chen_2023_adscft} experimentally studies AdS/CFT correspondence, confirming the EE for entanglement wedge as subsystem of bulk weakly-coupled scalar field is consistent with the RT formula~\cite{RT}. Inspired by the experimental progress, the experimental simulation of Gaussian fermionic field theory to study EE is worthy of further studying. 
Additionally, numerical studies of spin models also point out the non-trivial behaviors of correlation functions and entanglement of spin models on hyperbolic lattice~\cite{Gendiar_2020_area,Nishino_2024_holographic,2024arXiv240917235S}. 
Based on our numerical results and these works, along with the theoretical research on the relationship between EE and AdS space geometry, the investigation within the framework of field theory and holography to understand the area-law EE of ground states for free-fermion systems on hyperbolic lattice is an interesting topic for future work~\cite{witten,Klebanov1998,RT,maldacenaLarge1998,Viswanathan1998a,Viswanathan1998b,Henningson_1998}.

As it is feasible to simulate entanglement experimentally~\cite{Lin2024} while the experimental simulation of hyperbolic lattice has been achieved~\cite{Kollr2019,Yu_2020_T,Zhang2022,Lenggenhager2022,chen_2023_adscft,Huang2024}, this may provide us with a novel approach to study the geometry of the quantum system through entanglement. 
Furthermore, the area law of EE in both gapped and gapless systems suggests that it is efficient to study correlated systems on hyperbolic lattices with gapless emergent fermions by tensor-network-type numerical techniques~\cite{ORUS_2014,Cirac_2021_M,Xiang_2023}. We hope that our work can provide some inspiration to related fields in the future. Another interesting future direction is to study entanglement of non-Hermitian systems~\cite{Herviou_2019_E,Chang_2020_E,Chen_2021_nonhermi,Lee_2022,Modak_2021,Guo_2021,Tu_2022,Ortega_2022,Chen_2022_nonhermi,Kawabata_2023,Zou2023,Gal_2023,Hsieh_2023,ywz_2023,zlw_2024_E1,lsz_2024_E,zlw_2024_E2,2024arXiv240303259X,shi_2024,2024arXiv240613087M,2024arXiv240615564Y} on hyperbolic lattice, which is much more practical in, e.g., phononic systems where gain and loss are natural.

\begin{acknowledgments}
This work was in part supported by National Natural Science Foundation of China (NSFC) Grant No.~12474149 and No.~12074438. The calculations reported were performed on resources provided by the Guangdong Provincial Key Laboratory of Magnetoelectric Physics and Devices, No. 2022B1212010008.
\end{acknowledgments}

\appendix

\section{Hyperbolic lattice}
\label{app:lattice}
In this section we give details of constructing hyperbolic lattices discussed in Sec.~\ref{sec:fundamentals}. Additionally, we discuss the geometrical properties of hyperbolic lattice as well as the volume law of EE.

\subsection{Vertex inflation method of generating hyperbolic lattice}
\label{app:inflation}
 
The vertex inflation method or vertex-inflation tiling procedure for generating hyperbolic lattice was first purposed in the field of hyperbolic tensor-network theory~\cite{Boyle_2020_C,Jahn_2020_C} and then optimized for study in lattice many-body models~\cite{Chen_2023_S}. Here, we introduce our lattice set-up based on this method.

To start with, we generate a regular $p$-edges polygon at the center of the Poincar\'{e} disk and denote it as the $1$-st \textbf{ring} of the lattice. We then attach new sites to the $1$-st ring to form a new ring, and iteratively repeat this procedure. This finite-size lattices, named as flakes, can be divided into rings in order and every regular $p$-edges polygon is denoted as a \textbf{tile}. 
For every vertex of a tile, the vertex is affiliated to this tile. 
If a vertex doesn't have $q$ affiliated tile it is an \textbf{open vertex}. 
A vertex with $q$ neighboring vertices does not equal to not open since it may have less than $q$ affiliated tiles. If an open vertex has an nearest-neighboring vertex which is also open, the edge linking them is an \textbf{open edge}. The lattice set-up procedure is summarized as follows:
\begin{enumerate}
	\item For a $\{p,q,n\}$ lattice, we find all open vertices and their corresponding open edge on its outermost $n$-th ring. A vertex on $n$-th ring can either have zero or two open edge of which it is an endpoint.
	\item For every open vertex $i$ and one of its open edge, if its number of affiliated tiles is less than $q-1$, we identify the tile to which the open edge belongs and invert this tile. This process creates a new tile and an new open edge of which $i$ is an endpoint. 
	\item Otherwise, for every open vertex with $q-1$ affiliated tiles, we identify both two open edges it belongs to and generate a new tile based on them. 
	\item Go back to step one and repeat the whole process until all vertices on the $n$-th ring are no longer open. So far we have constructed a new ring and $\{p,q,n+1\}$ lattice.
\end{enumerate}

By using the above method, we can construct the entire lattice ring by ring. The procedure can be visualized as Fig.~\ref{fig:lattice}. The finite lattice generated by this method do not have dangling sites on the inner rings and it is natural to define the outermost ring as the boundary.

\subsection{Exponential growth of the size of the hyperbolic lattice}
\label{app:proof}
In this section, we give a brief proof of the exponential growth of the size of hyperbolic lattice. We start by considering $\{p,q\}$ lattice with $p\ge 4$ and $q\ge 5$. The proofs for the remaining cases are similar to the following proof.

For a $\{p,q,n-1\}$ lattice , all vertices on the outermost ($n-1$)-th ring can have either $2$ or $3$ nearest neighboring vertices to which is connected by an edge. We denote $N_n$ as the number of vertices on the $n$-th ring. The number of vertices having $2$ nearest neighboring vertices on the outermost ring is denoted as $N_{n-1,2}$, and the number of vertices having $3$ nearest neighboring vertices on the outermost ring is denoted as $N_{n-1,3}$ by analogy. Thus we have:
\begin{align}
	N_{n-1} = N_{n-1,2} + N_{n-1,3}
\end{align}
for any $n\ge 2$.

In the procedure of generating the lattice, the construction of $n$-th ring is only dependent on the ($n-1$)-th ring. 
Every $2$-neighboring vertex on the ($n-1$)-th ring directly has $q-2$ neighboring vertices on the $n$-th ring, and these $q-2$ vertices form $q-3$ tiles which need $p-3$ new vertices each.
Similarly, every $3$-neighboring vertex on the ($n-1$)-th ring directly has $q-3$ neighboring vertices on the $n$-th ring. These new neighboring vertices form $q-3$ tiles which need $p-3$ new vertices each. 

Besides, the edges on the ($n-1$)-th ring, whose number is equal to $N_{n-1}$, form $N_{n-1}$ tiles, each of which requires $p-4$ new vertices. Summarizing the above constraints, we have:
\begin{align}
	N_n=&(p-4)N_{n-1} + (q-2)N_{n-1,2} + (p-3)(q-3)N_{n-1,2} \nonumber\\
	&+(q-3)N_{n-1,3} + (p-3)(q-4)N_{n-1,3}\,. 
\end{align}
We also notice that each $3$-neighboring vertex on the ($n-1$)-th ring is directly connected to a vertex on ($n-2$)-th ring. That is:
\begin{align}
	N_{n-1,3} = (q-3)N_{n-2,3} + (q-2)N_{n-2,2} \nonumber
\end{align}
for any $n\ge 3$. And for $2$-neighboring vertex, the case is:
\begin{align}
	N_{n-1,2} =&(p-4)N_{n-2} + (p-3)(q-3)N_{n-2,2}  \nonumber\\
	&+ (p-3)(q-4)N_{n-2,3}\,. \nonumber
\end{align}
Summarizing the above results, we get the recursive relation:
\begin{align}
	N_n = (pq-2p-2q+2)N_{n-1} -N_{n-2}\,.
\end{align}

Solving this relation is equivalent to find the root of quadratic equation
\begin{align}
	x^2 - (pq-2p-2q+2)x + 1 =0
\end{align}
for $x$. As we directly have $N_1 = p$ and $N_2 = p^2q-2pq-2p^2+3p$, by solving above equation we find the formula of $N_n$, as 
\begin{align}
	N_n=&\frac{p}{2^{n+1}} \left( -1-\sqrt{\frac{t}{t-4}} \right) \left( t-2-\sqrt{t\left( t-4 \right)} \right) ^n \nonumber\\
	&+\frac{p}{2^{n+1}} \left( -1+\sqrt{\frac{t}{t-4}} \right) \left( t-2+\sqrt{t\left( t-4 \right)} \right) ^n \,,  
\end{align}
where $t=(p-2)(q-2)>4$. Finally summing over all the rings yields exponentially growing size of $\{p,q,n\}$ lattice:
\begin{align}
	N \sim \lambda^n \,,
\end{align}
where $\lambda$ depends on specific $p,q$ and can be analytically calculated. 

This shows an exponential growth of lattice size which is absolutely different from Euclidean case since Euclidean lattice grows as $N\sim n^2$. Some lattices are shown in Table.~\ref{tab:rings}, from which we can see the difference between hyperbolic case and Euclidean case.
\begin{table}[b]
	\caption{\label{tab:rings}
		Lattices construction by rings. This table shows the total number of sites on the $n$-th ring of different lattices.}
	\begin{ruledtabular}
		\begin{tabular}{ccccccc}
			lattice &1st&2nd&3rd&4th
			&5th&6th\\
			\hline
			$\{3,7\}$& 3 & 12 & 33 &87
			& 228 & 597 \\
			$\{4,5\}$& 4 & 20 & 76 &284
			& 1060 & 3956 \\
			$\{6,4\}$& 6 & 42 & 246 &1434
			& 8358 & 48714 \\
			$\{8,3\}$& 8 & 40 & 152 &568
			& 2120 & 7912 \\
			$\{8,8\}$& 8 & 280 & 9512 &323128
			& - & - \\
			\hline
			lattice &1st&5th&10th&100th
			&-&-\\
			\hline
			$\{3,6\}$& 3 & 27 & 57 &597 &  &  \\
			$\{4,4\}$& 4 & 36 & 76 &796 &  &  \\
			$\{6,3\}$& 6 & 54 & 114 &1194 &  &  \\
		\end{tabular}
	\end{ruledtabular}
\end{table}

\subsection{Entanglement volume law on hyperbolic lattice\label{appsec:volume}}
In this section, we show that $L_A$ of a subsystem approaches a finite fraction of the total number of lattice sites in the subsystem $N(V_{A})$ in the thermodynamical limit $n\rightarrow \infty$, and discuss the volume law of EE on hyperbolic lattice.

Consider a $\{p,q,n\}$ lattice as a subsystem $A$ in a larger lattice. The $L_A$ defined in Sec.~\ref{subsec:lattice} can be expressed as:
\begin{equation}
	L_A = (q-2)N_{n,2} + (q-3)N_{n,3} \, .
\end{equation}
By solving for $N_{n,2}$ and $N_{n,3}$ similar to the approach in Appendix~\ref{app:proof}, we find that the leading-order terms of $N_{n,2}$ and $N_{n,3}$ coincide with those of $N_n$. 
Therefore, $L_A/N$ becomes a finite fraction for sufficiently large subsystems or in the thermodynamical limit $n\rightarrow \infty$, e.g., approximately $1.732$ for the $\{4,5\}$ lattice.

The area law indicates that the EE is proportional to the degrees of freedom $d N(L^{D-1}_{A})$ on the boundary of the subsystem, where $d$ is the local Hilbert space on a lattice site, $D$ is the spacial dimension and $N(L^{D-1}_{A})$ is the number of boundary lattice sites of the subsystem.
On Euclidean geometry, $N(L^{D-1}_{A})$ is proportional to the boundary area $L^{D-1}_{A}$, where $L_{A}$ is the linear size of the subsystem~\cite{Barthel_2006_E}.  Meanwhile,  $N(L^{D-1}_{A})$ is equal to the total number of lattice bonds connecting the lattice sites in two complementary subsystems. Therefore, the scaling $S_{A}\sim L^{D-1}_{A}$ is referred to as the area law. 
In our work presented here, we also adopt the total number of bonds as the linear size $L_{A}$, as clarified in Sec.~\ref{subsec:gapless_area_law}, and numerically find the scaling $S_A\sim L_A$ still holds for both gapless (with finite DOS) and gapped free-fermion systems on hyperbolic lattices. Hence, we refer to this scaling on hyperbolic lattices as the area law. 

The volume law indicates that the EE is proportional to the total degrees of freedoms $d N(V_{A})$ within the subsystem, where $N(V_{A})$ is the total number of lattice sites in the subsystem. In Euclidean geometry, $N(V_{A})$ is proportional to the volume $V_{A}\sim L^{D}_{A}$ of the subsystem. 
Therefore, the scaling $S_A\sim L^{D}_{A}$ is referred to as the volume law. 
On hyperbolic lattices, however, the area of a region can scale as a finite fraction of its volume in the asymptotic limit (i.e., for sufficiently large subsystems), leading to $V_A\sim L_A$ geometrically. 
Consequently, on hyperbolic lattices, the area law can also be interpreted as $S_A\sim L_A \sim V_A$, which may alternatively be referred to as the volume law. However, to ensure the definition of area law is consistent on both Euclidean and hyperbolic lattices, we still regard the scaling $S_A\sim L_A$ on hyperbolic lattices as the area law.

\section{Supplemental data of numerical computations of EE through \textit{partition} $\text{\romannumeral 1}$ and \textit{partition} $\text{\romannumeral 2}$ }
\label{app:fit}
As detailed in Sec.~\ref{sec:fundamentals}, when studying EE, we use some different partition methods to investigate how the EE varies with the boundary $L_A$ as the size of the subsystem changes. 
  
\begin{figure*}
	\includegraphics[width=\linewidth]{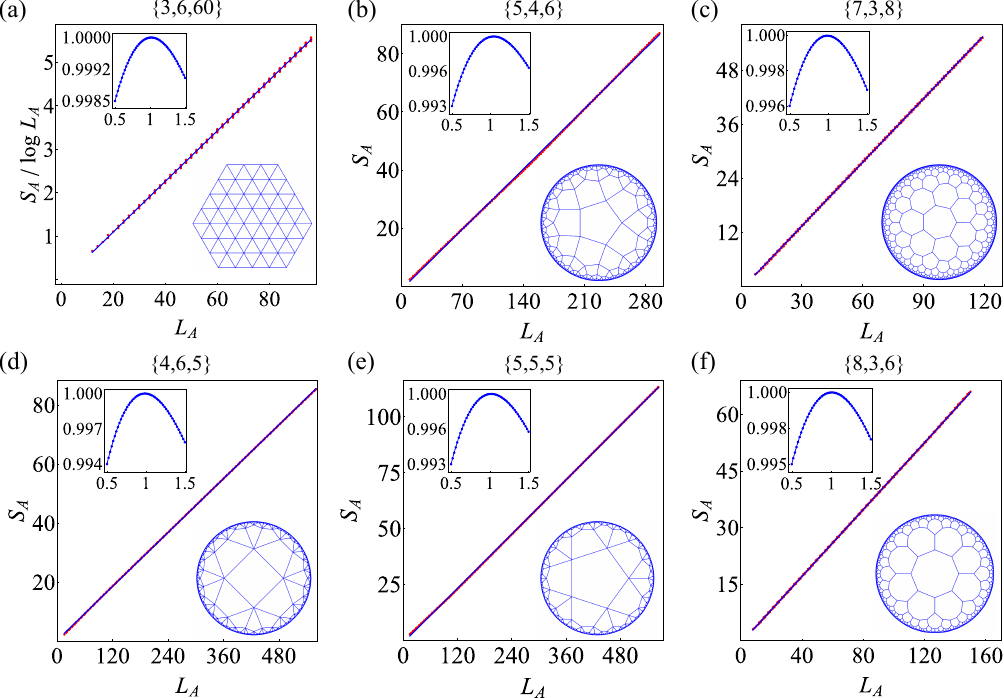}
 	\caption{\label{fig:app_partition1}Results of EE scaling fit for Hamiltonian $H_1$ with $t=1$ and $\mu=0$. Subsystems are generated through \textit{partition} $\text{\romannumeral 1}$ for different lattices. On Euclidean lattice $\{3,6,60\}$ (10800 sites) (a), the inset shows $R^2$ as a function of $\alpha$ in the fitting function $S_A / \log L_A=cL_A^\alpha+d$, while $R^2$ as a function of $S_A=aL_A^\alpha+b$ in the remaining hyperbolic case $\{5,4,6\}$ (6750 sites) (b), $\{7,3,8\}$ (15435 sites) (c), $\{4,6,5\}$ (6724 sites) (d), $\{5,5,5\}$ (15125 sites) (e) and $\{8,3,6\}$ (10800 sites) (f). All hyperbolic cases correspond with area law.}
\end{figure*}

The supplemental data of EE computed through \textit{partition} $\text{\romannumeral 1}$ with $t=1$ and $\mu=0$ for Hamiltonian $H_1$ on lattices different from those in the main text can be seen in Fig.~\ref{fig:app_partition1}. Here we anticipate the scaling function $S_A / \log L_A=cL_A^\alpha+d$ for Euclidean $\{3,6\}$ lattice which exhibit super-area law that can be seen in Fig.~\ref{fig:app_partition1}(a) while the hyperbolic cases all exhibit area law and we anticipate the scaling of EE is $S_A=a L_A^\alpha + b$. 
On Euclidean lattices, increasing subsystems size successively can result in many subsystems with different shapes and sizes sharing the same $L_A$, e.g., $\{3,6\}$ lattice in Fig.~\ref{fig:app_partition1}(a). In the main text the size of subsystem on Euclidean lattices grows discretely so that we have subsystems similar to the overall system. 
However, enlarging the size of the subsystem successively causes $L_A$ to increase successively in the hyperbolic case, as shown in Fig.~\ref{fig:app_partition1}(b-f). 
This enable us to study the growth of EE with the successively increasing boundary with numerous data, regardless of the exponential wall of the lattice size. Although this partitioning method may not maintain the symmetries, it still significantly distinguishes between area law and super-area law behavior of EE. 

\begin{figure*}
	\includegraphics[width=\linewidth]{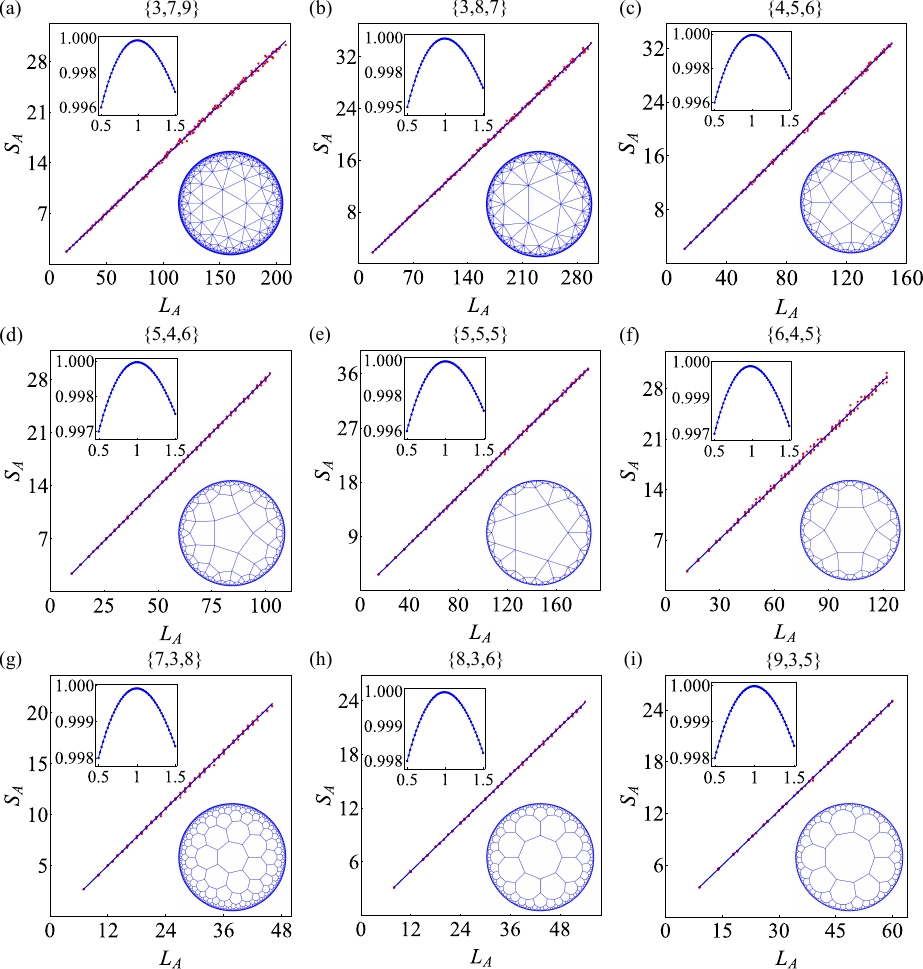}
	\caption{\label{fig:app_partition2}Results of EE scaling fit for Hamiltonian $H_1$ with $t=1$ and $\mu=0$, including $\{3,7,9\}$ (17328 sites) (a), $\{3,8,7\}$ (15123 sites) (b), $\{4,5,6\}$ (5400 sites) (c), $\{5,4,6\}$ (6750 sites) (d), $\{5,5,5\}$ (15125 sites) (e), $\{6,4,5\}$ (10086 sites) (f), $\{7,3,8\}$ (15435 sites) (g), $\{8,3,6\}$ (10800 sites) (h) and $\{9,3,5\}$ (7569 sites) (i) lattices. Subsystems are generated through \textit{partition} $\text{\romannumeral 2}$ for different hyperbolic lattices. The insets show $R^2$ as a function of $\alpha$ in the fitting function $S_A=aL_A^\alpha+b$.}
\end{figure*}

The results of EE computed through \textit{partition} $\text{\romannumeral 2}$ are exhibited in Fig.~\ref{fig:app_partition2}, where we use fitting function $S_A=a L_A^\alpha + b$. 
Because choosing subsystems too close to the boundary will cause finite-size effect, we define an internal region of the lattice, specify the size of subsystems and then randomly choose subsystems that can be composed of connected tiles. 
The results in Fig.~\ref{fig:app_partition2} are computed with Hamiltonian $H_1$ and we set $t=1$ and $\mu=0$, as are those shown in Fig.~\ref{fig:app_partition1}, and the blue lines show the fitting functions with $\alpha = 1$. 
Even with the same size or the same $L_A$, subsystems partitioned through this method can have various possible shapes and do not maintain the same symmetries. However, the symmetries of these subsystems do not affect the scaling behavior of EE. 
From the results, we find that linearity still demonstrates that the best description between EE and boundary is area law. 

\section{Numerical study of DOS}
\label{app:DOS}
Based on our considerations in the main text, we need to verify that the Hamiltonian $H_1$ is indeed gapless on lattices we considered. Because the geometric properties of hyperbolic lattice induce exotic behavior of free fermions, we use DOS as the verification.

Additionally, as aforementioned, the size of the system grows exponentially with $n$, resulting numerical difficulties in exact diagonalization (ED) approach. 
Therefore, we use the Haydock recursion method~\cite{Haydock1972,Haydock1973,Gaspard1973,Haydock1975,Haydock1984,Mosseri_2023_D} to acquire DOS in the thermodynamical limit.

\begin{figure*}
	\includegraphics[width=\linewidth]{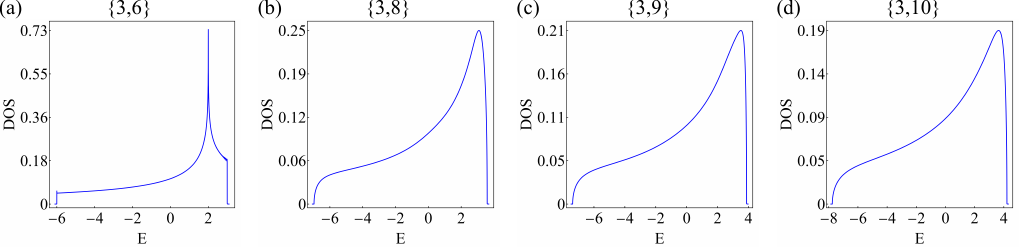}
	\caption{\label{fig:haydock_dos}Normalized DOS computed by the Haydock recursion method for Hamiltonian $H_1$ with $t=1$ and $\mu=0$. The lattices chosen here share the same $p=3$ with $q=6,8,9,10$, while $\{3,7\}$ has been shown in Fig.~\ref{fig:gapless_entropy}(b2). Through this method we verify that these systems are indeed gapless.}
\end{figure*}

\subsection{Haydock recursion approach to DOS}
We can calculate local density-of-states (LDOS) at a particular site $j$ by Green's function:
\begin{align}
	\label{ldos_app}
	\rho_j(E) = -\lim_{\epsilon\rightarrow 0^+} \frac{1}{\pi}\mathrm{Im}\left< j \lvert G(E+\mathrm{i}\epsilon) \rvert j \right>\,.
\end{align}
The Green's function $G_{ij}(E)=\left< i \lvert (E-H)^{-1} \rvert j \right>$ can be decomposed into contributions from moments of the Hamiltonian $G_{ij}(E)=E^{-1} \left< i \lvert 1+\sum_{n}H^n/E^n \rvert j \right>$.

The Haydock recursion method~\cite{Haydock1972,Haydock1973,Haydock1975}, also known as the continued-fraction method, give a method to compute the diagonal matrix element of $G(E)$:
\begin{align}
	\label{recursiverelation}
	G_{jj}(E)&=\left< l_1 \lvert G(E) \rvert l_1 \right> \nonumber\\
    &=\frac{1}{E-a_1-\frac{b_{1}^{2}}{E-a_2-\frac{b_{2}^{2}}{\cdots}}}\,.
\end{align}
Here $\left| l_1 \right>$ is a unit vector that has non-zero component at site $j$ only. The rational continued-fraction coefficients $a_i$ and $b_i$ in Eq.~(\ref{recursiverelation}) can be obtained by the following recursive relation:
\begin{align}
	\begin{cases}
		a_i = \left< l_i \lvert H \rvert l_i \right>
		\\ \left| n_{i+1} \right>  = (H-a_i) \left| l_i \right> - b_{i-1}\left| l_{i-1} \right>
		\\ b_{i}=\sqrt{\left< n_{i+1}\lvert n_{i+1}\right>}
		\\ \left| l_{i+1} \right> = \frac{1}{b_{i}} \left| n_{i+1} \right>
	\end{cases},
\end{align}
where $i=1,2,3\dots$ and $b_0=0$. For gapless systems, the coefficients $a_i$ and $b_i$ converge to the asymptotic value $a_\infty$ and $b_\infty$ for sufficiently large lattices and give the band edges:
\begin{align}
	\label{gapless_edges}
	E_\pm=a_\infty\pm2b_\infty\,.
\end{align}
For gapped systems with single band gap, which is the case of Hamiltonian $H_2$, the coefficients $b_i$ converges to two asymptotic limit $\overline{b}$ and $\underline{b}$ when
 $n\rightarrow\infty$~\cite{Gaspard1973}:
 \begin{align}
 	\label{gapped_edges}
 	E_+-E_-&=2(\overline{b}+2\underline{b}) \nonumber\\
 	\Delta&=2(\overline{b}-2\underline{b}) \,,
 \end{align}
where $\Delta$ is band gap.

To accurately compute the rational coefficients $a_n$ and $b_n$ to the order $n$, the shortest graphic path from site $j$ to boundary $R_j$ as defined in Sec.~\ref{sec:fundamentals} should be at least $n$. 
Then we introduce a proper fraction termination: 
\begin{align}
	t\left( E \right) =\frac{E-a_{\infty}-\sqrt{\left( E-a_{\infty} \right) ^2-4b_{\infty}^2}}{2b_{\infty}^2}
\end{align}
for Hamiltonian $H_1$, where $a_{\infty}$ and $b_{\infty}$ are chosen as the converged $a_n$ and $b_n$ for large $n$. In gapped system the fraction termination can be more complicated~\cite{Gaspard1973,Haydock1984}, for Hamiltonian $H_2$ we use:
\begin{align}
	t\left( E \right) =\frac{\left( E-A \right) ^2+A^2-B+2b_{\infty}^2-X\left( E \right)}{2b_{\infty}^2\left[ \left( E-A \right) +\left( a_{\infty}-A \right) \right]}\,,
\end{align}
where $A=\frac{1}{4}\sum{t_i}$, $B=\frac{1}{4}\sum{t_{i}^{2}}$, $X=\prod{\sqrt{E-t_{i}^{2}}}$ and $t_i,i=1\dots4$ are band edges which can be obtained by the asymptotic coefficient in Eq.~\ref{gapped_edges}.

After deciding the termination, the LDOS at site $j$ is given by Eq.~(\ref{ldos_app}) and Eq.~(\ref{recursiverelation}). 
Since for regular tillings sites in the bulk are all equivalent if the lattice is sufficiently large, the LDOS is DOS up to a normalization factor~\cite{Mosseri_2023_D}.

\subsection{Numerical results of DOS}
We show some results of DOS which are computed on different lattices with up to $10^7$ sites for Hamiltonian $H_1$ in Fig.~\ref{fig:haydock_dos}. Here we compute DOS on $p=3$ lattices and verify that they are gapless. Since this method's memory consumption scales linearly with the lattice size, it significantly exceeds the computational limits of ED methods. This method can be applied to arbitrarily large lattice (one can obtain result for lattice size up to $10^9$ sites~\cite{Mosseri_2023_D}) but our results here are sufficient to determine whether the system is gapped or gapless in the thermodynamical limit. 

We notice that the thermodynamical DOS obtained through this method is different from that computed on finite lattices through ED, indicating that the computation of EE may exhibit finite-size effect. 

\begin{figure}
	\includegraphics[width=\columnwidth]{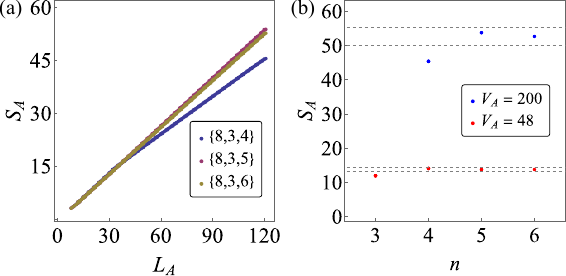}
	\caption{\label{fig:finitesize} Finite-size scaling analysis of EE by the example of $\{8,3\}$ lattice. (a) EE for the same subsystems computed on $\{8,3,n\}$ lattice with $n=4,5,6$. (b) EE for two specific subsystems ($V_A = 48$ and $200$ sites) computed on $\{8,3,n\}$ lattice, where dashed lines show $\pm5\%$ intervals of EE computed on $\{8,3,6\}$ lattice. Results are computed for Hamiltonian $H_1$ with $t=1$ and $\mu=0$.}
\end{figure}

\begin{table}[b]
	\caption{\label{tab:superareaanalysis}
	Results of adjusted coefficient of determination $\bar{R}^2$ and dimensionless error estimations, including mean absolute percentage error (MAPE) and relative absolute error (RAE) for EE data shown in Fig.~(\ref{fig:gapless_entropy}) are presented. In Euclidean case, the super-area law $S_A / \log L_A=a_2L_A+b_2$ is better fit, while in hyperbolic case the area law $S_A = a_1L_A+b_1$ is better.}
	\begin{ruledtabular}
		\begin{tabular}{c|c|c|c|c}
			lattice                       & fit function & $\bar{R}^2$ & MAPE ($\%$)    & RAE                     \\ \hline
			\multirow{2}{*}{$\{4,4,40\}$} & area         & 0.998466    & 3.66006  & $3.69134\times 10^{-2}$ \\ 
										  & super-area   & 0.999929    & 0.338284 & $7.64020\times 10^{-3}$ \\ \hline
			\multirow{2}{*}{$\{3,7,9\}$}  & area         & 0.999962    & 0.495449 & $5.79884\times 10^{-3}$ \\ 
										  & super-area   & 0.998026    & 3.12212  & $4.06104\times 10^{-2}$ \\ \hline
			\multirow{2}{*}{$\{4,5,6\}$}  & area         & 0.999918    & 0.832573 & $8.48647\times 10^{-3}$ \\ 
										  & super-area   & 0.998041    & 2.55588  & $4.20934\times 10^{-2}$ \\
			\end{tabular}
	\end{ruledtabular}
\end{table}

\begin{figure}
	\includegraphics[width=\columnwidth]{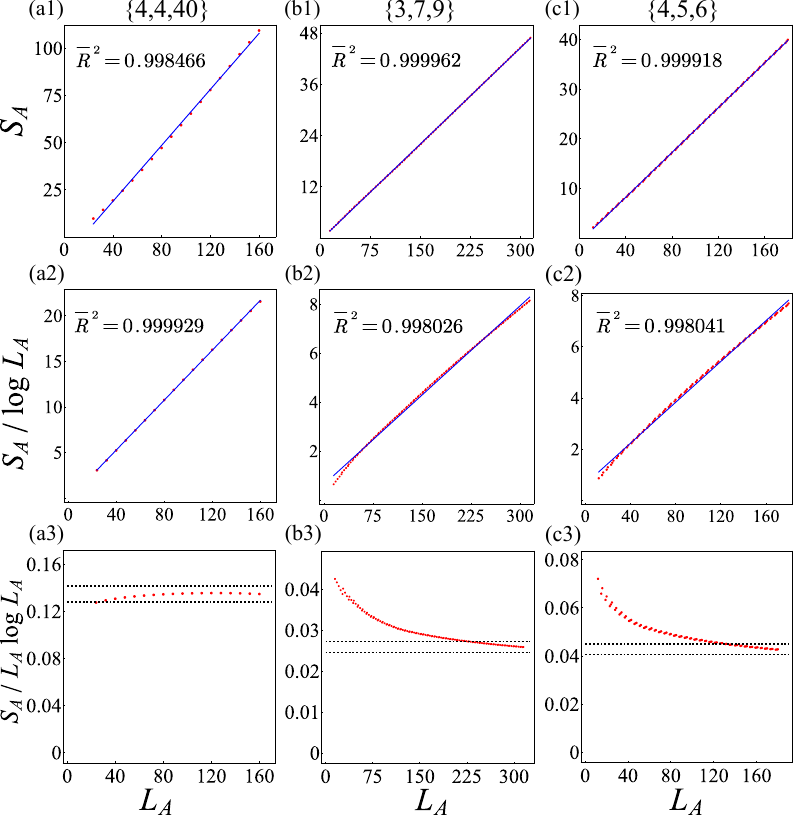}
	\caption{\label{fig:superareaanalysis} Column (a-c) correspond to the EE data of $\{4,4,40\}$, $\{3,7,9\}$, $\{4,5,6\}$ shown in Fig.~(\ref{fig:gapless_entropy}). In row (1-2), we present the linear fits $S_A = a_1L_A+b_1$ and $S_A / \log L_A=a_2L_A+b_2$, along with the adjusted coefficient of determination $\bar{R}^2$. The blue lines represent the respective linear fit functions. In row (3), we plot $S_A / L_A\log L_A$ as a function of $L_A$, where dashed lines represent $\pm5\%$ intervals of the data point corresponding to the biggest $L_A$.}
\end{figure}

\section{Finite-size scaling analysis of EE}
\label{app:finitesizescaling}
As we focus on EE for subsystems in the bulk, in this section we perform the finite-size scaling analysis to demonstrate that the boundary effect is considered in our numerical computations. We take the $\{8,3\}$ lattice for Hamiltonian $H_1$ with $t=1$ and $\mu=0$ as an example. In Fig.~\ref{fig:finitesize}(a), we show the EE for the same subsystems (\textit{partition} $\text{\romannumeral 1}$ is taken) computed on $\{8,3,4\}$ ($768$ sites), $\{8,3,5\}$ ($2888$ sites) and $\{8,3,6\}$ ($10800$ sites) lattice respectively, where the largest subsystem studied here is identical to $\{8,3,3\}$ lattice ($V_A = 200$ sites). The optimal fit is obtained when $\alpha$ in the fitting function $S_A=aL_A^\alpha+b$ equals to $0.859$, $1.010$ and $0.997$ for $\{8,3,4\}$, $\{8,3,5\}$ and $\{8,3,6\}$ lattice respectively. 
In Fig.~\ref{fig:finitesize}(b), two specific subsystems, identical to $\{8,3,2\}$ ($V_A = 48$ sites) and $\{8,3,3\}$ ($V_A = 200$ sites) lattice respectively, are chosen to compute EE on $\{8,3,n\}$ lattices with $n=3,4,5,6$. Notably, the $\{8,3,3\}$ subsystem is not computed on $\{8,3,3\}$ lattice itself. The gray dashed lines represent $\pm 5\%$ intervals of the EE computed on $\{8,3,6\}$ lattice. 
From these results, we observe that by minimizing boundary effects through enlarging the lattice, we obtain a linear fit of the EE for subsystems in the bulk. 
Through the finite-size scaling analysis, we show that the boundary effect is considerably excluded to obtain the EE for subsystems in the bulk of the hyperbolic lattice.

\section{Numerical analysis of super-area law scaling}
\label{app:superarea}
In this section, we present the numerical analysis to show that the EE does not scale as super-area law scaling. Specifically, we take the EE data for the gapless free fermions with finite DOS presented in Fig.~(\ref{fig:gapless_entropy}), Sec.~\ref{subsec:gapless_area_law} as an example. 
We anticipate the scaling functions $S_A = a_1L_A+b_1$ and $S_A / \log L_A=a_2L_A+b_2$ and fit the EE data. 
As logarithms can be hard to detect, we select several dimensionless evaluation metrics to compare these two fits, including the adjusted coefficient of determination $\bar{R}^2$, mean absolute percentage error (MAPE) given by $\frac{100\%}{N}\sum_i^N{\left| \frac{\hat{y}_i-y_i}{y_i} \right|}$, and relative absolute error (RAE) given by $\frac{\sum_i^N{\left| y_i-\hat{y}_i \right|}}{\sum_i^N{\left| y_i-\overline{y} \right|}}$. 
Here $N$ is the total number of data points, $y_i$ denotes the true values, $\overline{y}$ denotes the mean of $y$ and $\hat{y}_i$ denotes the predicted values. A higher $\bar{R}^2$ approaching $1$ and lower error estimations approaching $0$ indicate a better fitting.
The results of these evaluations are listed in Table.~(\ref{tab:superareaanalysis}), where one can find that area-law fitting is better in hyperbolic case. 
We also visualize the fittings in row (1-2) of Fig.~(\ref{fig:superareaanalysis}). In row (3), we plot $S_A / L_A\log L_A$ as a function of $L_A$ which only converge in the Euclidean case.
Therefore, we conclude that the EE does not scale as super-area law.

\begin{figure}
	\includegraphics[width=\columnwidth]{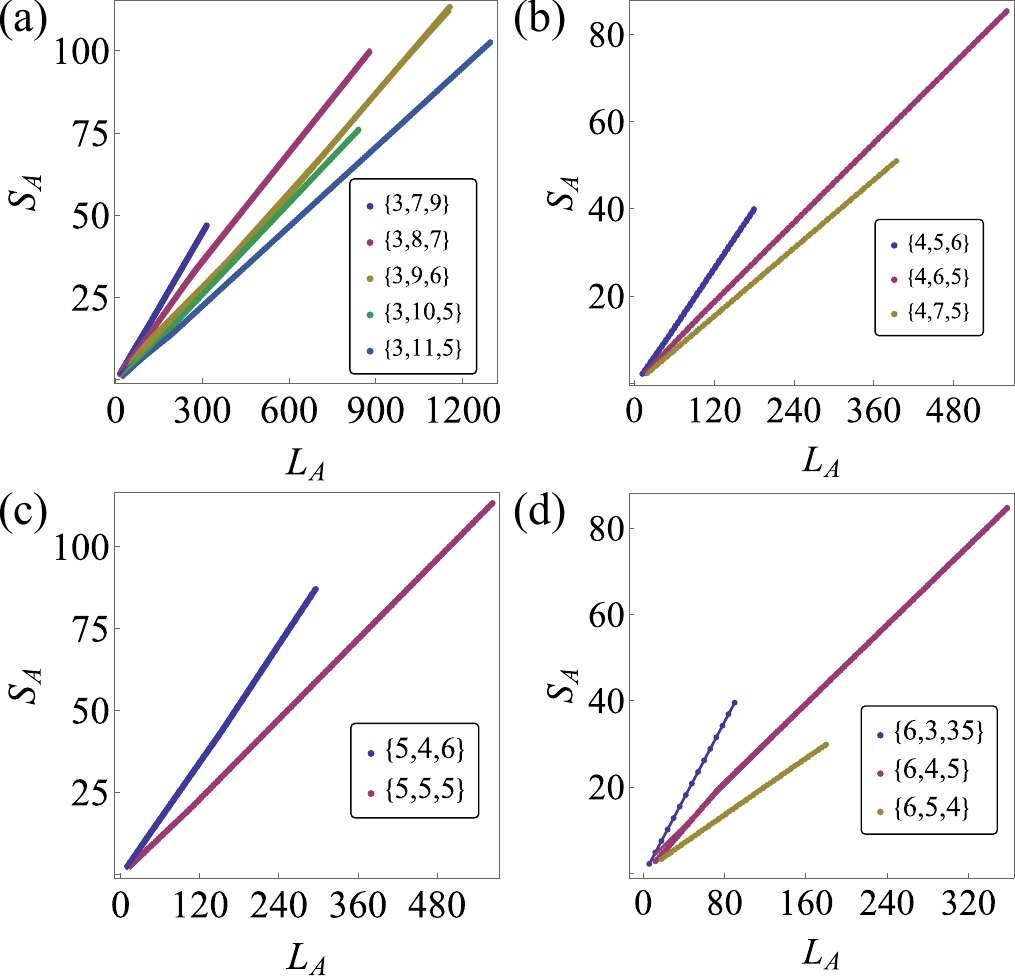}
	\caption{\label{fig:asymptotic_behavior} Asymptotic behavior of coefficient $a$. The cases for $p=3,4,5$ and $6$ all indicate that when $q$ increases, $a$ decreases. Results are computed for Hamiltonian $H_1$ with $t=1$ and $\mu=0$.}
\end{figure}

\section{Asymptotic behavior of scaling coefficient of area law}
\label{app:asymtotic}
In this section, we study how EE varies with $q$ when $p$ is fixed.
The number of nearest neighboring sites of a given site on hyperbolic lattice, labeled as $q$ as aforementioned, can increase successively. 
From our findings in the main text, EE is proportional to the boundary of subsystem $L_A$, which is a function of $q$, thus the area-law scaling coefficient $a$ should also be related to $q$. 
We study EE for Hamiltonian $H_1$ on $p = 3,4,5$ and $6$ hyperbolic lattices with successively increased $q$ and the results are shown in Fig.~\ref{fig:asymptotic_behavior}. The results all indicate that as the number of adjacent sites per site $q$ increases, the coefficient $a$ decreases. 

Due to computational difficulties on hyperbolic lattices, such as the exponentially growing lattice size and finite-size effect, it is hard to perform the scaling analysis for lattice with larger $q$. However, our results here indicate a monotonically decreasing relationship between $q$ and $a$.
It makes sense to explore the relationship of $a$ as $q$ increases, as this may reveal the asymptotic behavior of EE and provide us an new insights into the hyperbolic geometry.



\providecommand{\noopsort}[1]{}\providecommand{\singleletter}[1]{#1}%

\end{document}